\pdfoutput=1
\documentclass[]{aa}
\usepackage{graphicx}
\usepackage{amsmath}
\usepackage{latexsym}
\usepackage{longtable}
\usepackage{natbib}
\usepackage[latin1]{inputenc}
\usepackage{graphics}
\usepackage{epsfig}
\usepackage{graphicx}

\begin{document}

\title{Kolmogorov cosmic microwave background sky}
\author{V.G.Gurzadyan\inst{1}, A.E.Allahverdyan\inst{1}, T.Ghahramanyan\inst{1}
A.L.Kashin\inst{1}, H.G.Khachatryan\inst{1}, A.A.Kocharyan\inst{1,2},
H.Kuloghlian\inst{1}, S.Mirzoyan\inst{1}, E.Poghosian\inst{1}, G.Yegorian\inst{1}}

\institute
{\inst{1} Yerevan Physics Institute and Yerevan State University, Yerevan,
Armenia\\
\inst{2} School of Mathematical Sciences, Monash University, Clayton, Australia
}

\date{Received (\today)}

\titlerunning{CMB}

\authorrunning{V.G.Gurzadyan et al}

\abstract{
A new map of the sky representing the degree of randomness in the cosmic microwave background (CMB) temperature has been obtained. The map based on estimation of the Kolmogorov stochasticity parameter clearly distinguishes the contribution of the Galactic disk from the CMB and reveals regions of various degrees of randomness that can reflect the properties of inhomogeneities in the Universe. For example, among the high randomness regions is the southern non-Gaussian anomaly, the Cold Spot, with a stratification expected for the voids. Existence of its counterpart, a Northern Cold Spot with almost identical randomness properties among other low-temperature regions is revealed.  By its informative power, Kolmogorov's map can be complementary to the CMB temperature and polarization sky maps.
}


\maketitle

\section{Introduction}

Since the discovery of the anisotropy of the cosmic microwave background (CMB) radiation by COBE, the pixelized sky maps of CMB temperature were among the basic sources of cosmological information \cite{COBE}. Later experiments, including another full sky coverage by Wilkinson Microwave Anisotropy Probe (WMAP) \cite{WMAP_t}, produced higher sensitivity and resolution maps, enabling determination of the values of a set of cosmological parameters. A new channel of information open the CMB polarization measurements \cite{DASI,WMAP_p}, and higher accuracy data are expected in forthcoming experiments.
Along with the main cosmological parameters, the CMB maps have revealed tinier, non-Gaussian features (see Eriksen et al. 2004; Copi et al. 2007, 2008; Gurzadyan et al. 1997, 2005, 2007, 2008; Hansen et al. 2008) which would also be dealt with by more accurate measurements.     

Here we advance another concept of a digital sky, namely, a map of the degree of CMB temperature randomness using the Kolmogorov stochasticity parameter (KSP). The stochastic parameter and the statistic introduced by Kolmogorov \cite{K} lead to measurement of the objective randomness degree of finite sequences resulting from dynamical systems or number theory (Arnold 2008).

The creation of Kolmogorov's map (K-map) estimates the stochasticity parameter for ordered sequences of the CMB temperature values assigned to a set of pixels in a given region of the sky, as it was first performed in Gurzadyan \& Kocharyan (2008). In the analysis below we use the maps obtained by the WMAP \cite{WMAP_t}. While the resolution of stochasticity parameter mapping is based on the pixel sequences, higher resolution data expected in experiments will produce more informative K-maps than the one based on the currently available WMAP data.

It does not seems trivial that the resulting digitized sky map would dublicate or, on the contrary, differ from the features in the conventional temperature maps. This concerns, for example, non-Gaussian anomalies such as the Cold Spot \cite{c_spot,Cruz}, the nature of which is associated with underdense regions, the voids \cite{void,MN}. The voids are shown to act as hyperbolic lenses \cite{GK3} in a Friedmann-Robertson-Walker universe with perturbations of the metric, i.e. able to induce loss of information \cite{Arnold} on the propagating photon beams, hence a randomization in the CMB temperature anisotropy distribution. The role of voids as diverging lenses has been noticed also Das \& Spergel (2008);  therefore, at least within these effects the K-map can reflect definite properties of the large-scale matter distribution in the Universe.

The first remarkable result in the obtained K-map is the clearly distinguished Galactic disk, with high and nearly uniform Kolmogorov's parameter, thus outlining it upon the CMB contribution in the rest of the sky.    
Then, the Cold Spot is shown to have not only high degree of randomness, but also a stratification expected for the voids (see \cite{DS}), namely the increase of the randomness by radius, i.e. towards the walls. Using the latter property as an indicator, its counterpart, a Northern Cold Spot, is identified among the other cold regions of the sky. Other notable conclusions are drawn, namely, by a specially chosen strategy of computations' invariant values (i.e. independent on the split) of degree of randomness are assigned to the hemispheres; also, the simulated Gaussian isotropic maps are shown to possess an essentially lower degree of randomness than the real CMB maps.

\section{Kolmogorov Stochasticity Parameter}

Kolmogorov's parameter is defined for $n$ independent values $\{X_1,X_2,\dots,X_n\}$ of the real-valued random variable $X$ ordered in an increasing manner $X_1\le X_2\le\dots\le X_n$. The cumulative distribution function  (CDF) of $X$ is defined as
$$
F(x) = P\{X\le x\},\ 
$$
and the {\it empirical} distribution function $F_n(x)$ as
\begin{eqnarray*}
F_n(x)=
\begin{cases}
0\ , & x<X_1\ ;\\
k/n\ , & X_k\le x<X_{k+1},\ \ k=1,2,\dots,n-1\ ;\\
1\ , & X_n\le x\ .
\end{cases}
\end{eqnarray*}
Kolmogorov's stochasticity parameter $\lambda_n$ is defined by \cite{K},\cite{A_KSP}
\begin{equation}\label{KSP}
\lambda_n=\sqrt{n}\ \sup_x|F_n(x)-F(x)|\ .
\end{equation}

In (Kolmogorov 1933), Kolmogorov proved\footnote{Arnold \cite{Arnold_ICTP} refers to it as ``the astonishing Kolmogorov's theorem".}
that for any continuous CDF $F$
$$
\lim_{n\to\infty}P\{\lambda_n\le\lambda\}=\Phi(\lambda)\ ,
$$
where $\Phi(0)=0$,
\begin{equation}
\Phi(\lambda)=\sum_{k=-\infty}^{+\infty}\ (-1)^k\ e^{-2k^2\lambda^2}\ ,\ \  \lambda>0\ ,\label{Phi}
\end{equation}
the convergence is uniform, and $\Phi$, the Kolmogorov's distribution, is independent of $F$.
The mean value of $\Phi(\lambda)$ corresponds to $\lambda \approx 0.87$.
KSP is applied to measure the objective stochasticity degree of datasets (Arnold 2008a, 2008b). 

\section{Analysis}

As in Gurzadyan \& Kocharyan (2008), we used the 5-year data of WMAP obtained in the W band (94 GHz), as well as the corresponding foreground reduced map. The W band has the best resolution and the least contamination by the Galactic synchrotron radiation. The negligible role of the Galactic dust emission for KSP was revealed in Gurzadyan \& Kocharyan (2008) by the foreground-reduced maps in the Q and V bands. HEALPix \cite{HP} representation of the temperature maps was applied to the dataset of temperatures assigned to the pixel sequences. 

In the calculations $F(x)$ was a CDF for the Gaussian distribution; first $\lambda_n^{(k)}$, then $\Phi(\lambda_n^{(k)})$ via Eq. (2) were obtained. 

\begin{figure}[ht]
\centerline{\epsfig{file=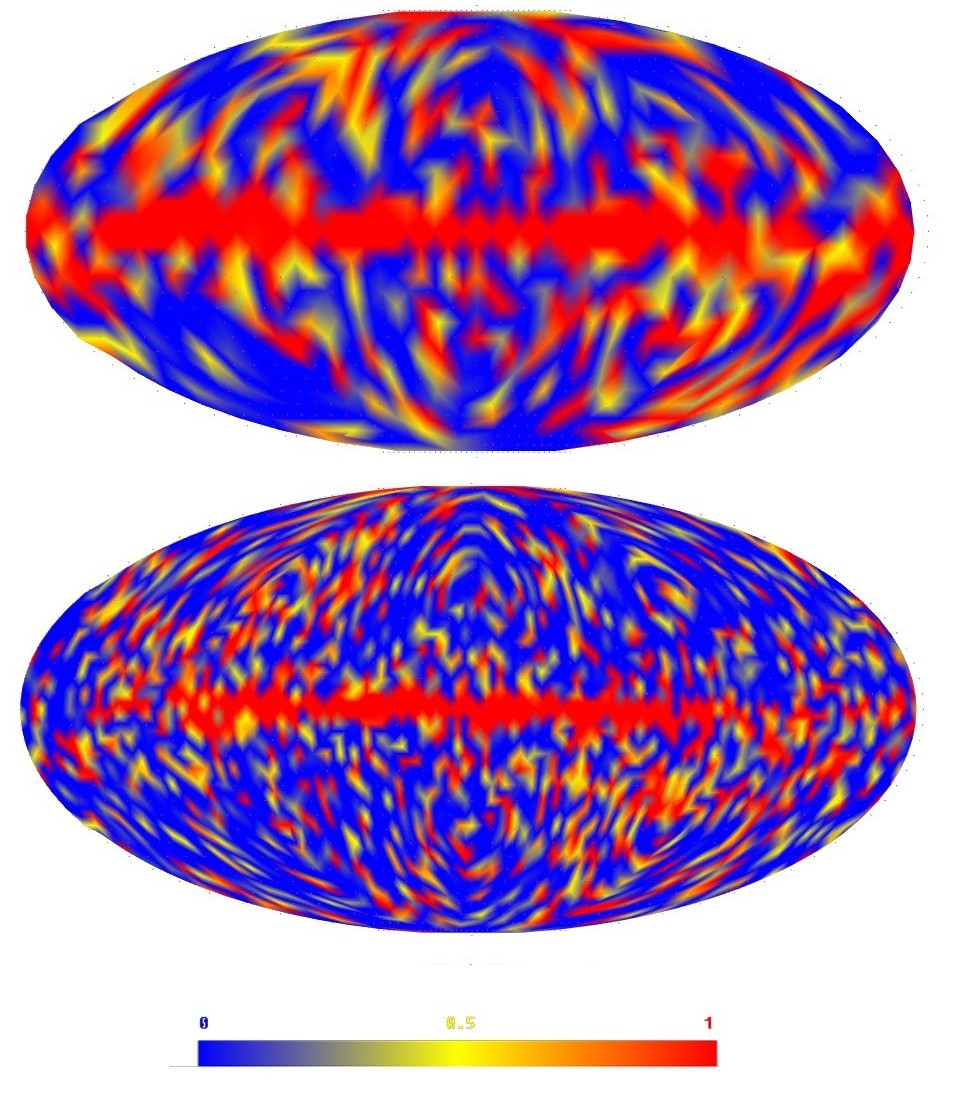,width=0.5\textwidth}} \vspace*{8pt}
\caption{Kolmogorov maps i.e. the degree of randomness in CMB sky. WMAP's 5-year W-band, 94 GHz data are used; upper map is for Nside=8, the lower for Nside=16. The Galactic disk is clearly distinguished.}
\end{figure}

\begin{figure}[ht]
\centerline{\epsfig{file=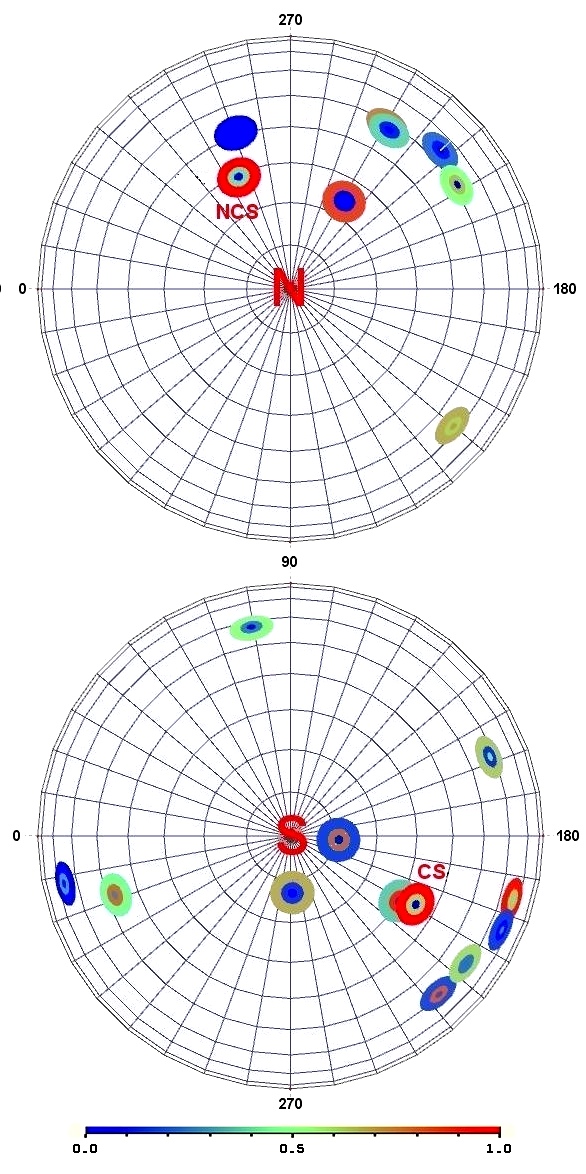,width=0.5\textwidth}} \vspace*{8pt}
\caption{Positions of the 20 cold spots, i.e. negative mean temperature regions, in the hemispheres. The Cold Spot (CS) and the Northern Cold Spot (NCS) are highlighted. The spots are stratified according to the randomness scale $\Phi$.}
\end{figure}

\begin{figure}[ht]
\centerline{\epsfig{file=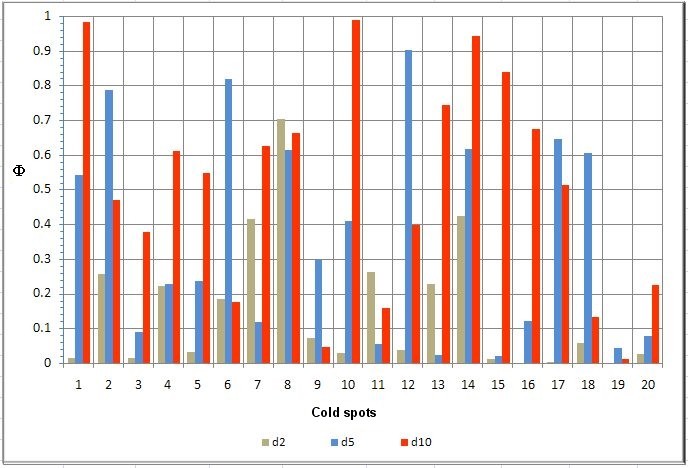,width=0.5\textwidth}} \vspace*{8pt}
\caption{The stratification of the degree of randomness of 20 cold spots within 2, 5 and 10$^{\circ}$ diameters, respectively. For voids the randomness is expected to increase towards larger radii, i.e. the walls; No.1 is the Cold Spot, No.10 is the Northern Cold Spot.}
\end{figure}

\begin{figure}[ht]
\centerline{\epsfig{file=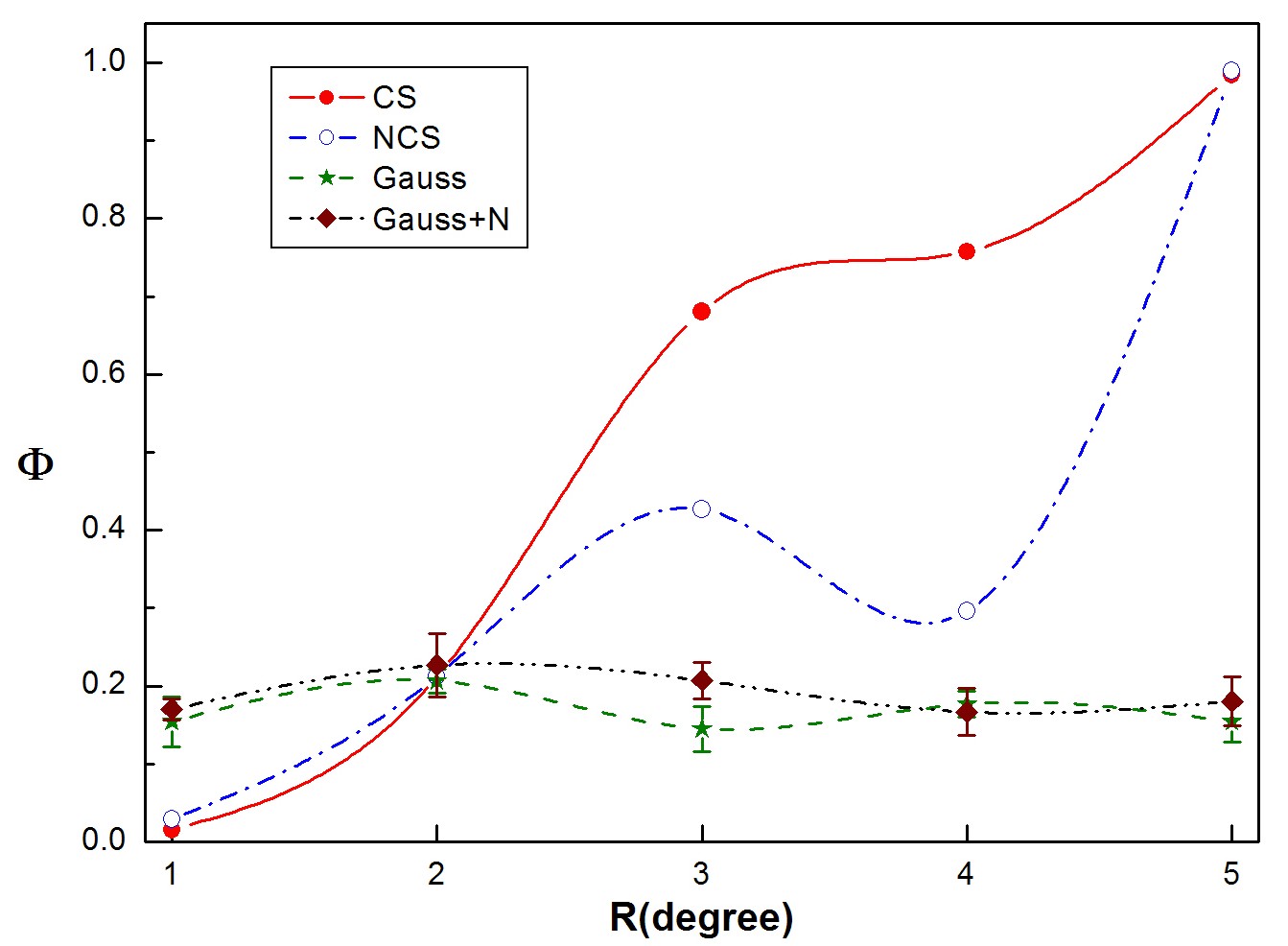,width=0.5\textwidth}} \vspace*{8pt}
\caption{The dependence of the degree of randomness $\Phi$ vs the radius for the CS and the NCS. The mean $\Phi$ for 20 same size spots of Gaussian maps with the WMAP's noise and for isotropic ones are also shown.}
\end{figure}

\begin{figure}[ht]
\centerline{\epsfig{file=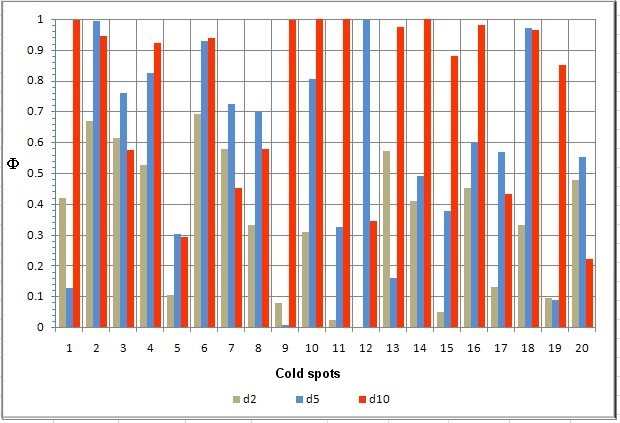,width=0.5\textwidth}} \vspace*{8pt}
\caption{The same as in Fig 3 but for the WMAP's map with removed first 20 multipoles.}
\end{figure}

Figure 1 represents full-sky K-maps at $Nside$ = 8 and 16, i.e. a resolution about 7$^{\circ}$.3 and 3$^{\circ}$.65, respectively, with the assigned color scale for $\Phi$.  
The Galactic disk is distinguished with almost constant $\Phi=1$, with slight variations only in the Galactic anticenter direction. 

Then, Kolmogorov's parameter was obtained by a split of the hemispheres, excluding the Galactic disk $|b|< 20^{\circ}$ belt, into several scale regions; namely, we covered the hemispheres with the regions of the following formats: 60$^{\circ}$ x 30$^{\circ}$, 60$^{\circ}$ x 24$^{\circ}$, 60$^{\circ}$ x 20$^{\circ}$; 30$^{\circ}$ x 30$^{\circ}$, 30$^{\circ}$ x 24$^{\circ}$, 30$^{\circ}$ x 20$^{\circ}$. Moreover, the partitions themselves were modified; i.e. the initial axis (meridian) at each split was shifted continuously, in order to exclude any dependence on the mode of split. The results demonstrated definite invariance to the mode of split (visible by the small variance), thereby enabling a degree of randomness to be assigning to the hemispheres, as represented in Table 1.

\begin{table}[ht]
\centering
\caption{Mean values and variances for $\Phi(\lambda)$ for the northern and southern Galactic hemispheres.}
\begin{tabular}{l c c}
\\
\hline\hline
Hemisphere        &  Mean($\Phi$) & Var($\Phi$)\\ [0.5ex]
\hline
Northern             & 0.53  & 0.002 \\
Southern             & 0.61  & 0.002 \\[1ex]
\hline
\end{tabular}
\end{table}

\begin{table}[ht]
\caption{The degree of randomness $\Phi$ for the CS and the NCS. The columns contain coordinates of their centers, the pixel counts within 1$^{\circ}$ ($N_1$), 3$^{\circ}$ ($N_3$), and 5$^{\circ}$($N_5$)
radii, and values for $\Phi$ for each region. Mean $\Phi$ for 20 simulated Gaussian maps with superposed WMAP's noise (Gauss+noise) and without noise, i.e. isotropic map.}

\centering
\begin{tabular}{c c c c c }
\\
\hline\hline
          &      Cold       & Northern        & Gauss +   &       \\
          &      Spot       & Cold Spot       & noise     & Gauss \\[0.5ex]
\hline
  $l$     & 208$^{\circ}$.7 & 294$^{\circ}$.8 &           &       \\       
  $b$     & -55$^{\circ}$.6 &  60$^{\circ}$.8 &           &       \\       
 $N_1$    &   239           &    243          &  239      & 239   \\       
 $N_2$    &  2155           &   2154          & 2157      & 2157  \\      
 $N_3$    &  5987           &   5989          & 5983      & 5984  \\      
 $\Phi_1$ &  0.02           &   0.03          & 0.17      & 0.15  \\      
 $\Phi_2$ &  0.68           &   0.43          & 0.15      & 0.15  \\      
 $\Phi_3$ &  0.98           &   0.98          & 0.18      & 0.15  \\      
\hline
\end{tabular}
\end{table}


A sample of 20 cold spots, i.e. of regions of 5$^{\circ}$ radius and with negative mean temperature, was revealed via the scan of the full sky, again excluding the Galactic disk (Fig. 2). Their stratification over $\Phi$ within the 1$^{\circ}$, 2$^{\circ}$.5, and 5$^{\circ}$ radii was performed, to test the increase in the randomness with radius as expected for the voids. Figure 3 exhibits the distribution of the randomness over the radii for the spots.

The spot of coordinates  $l$=294$^{\circ}$.8, $b$=60$^{\circ}$.8, denote it NCS (No.10 in Fig. 3), possesses properties similar to the CS (No.1) properties both by the value and radius dependence of the $\Phi$, see Fig. 4 and Table 2.  For comparison, the mean $\Phi$ for 20 simulated Gaussian maps with both superposed WMAP's noise (as in Gurzadyan \& Kocharyan (2008)) and isotropic maps are represented as well. The difference in the behaviors of the real spots, CS and NCS, and of the Gaussian ones with and without noise, is visible. 

To trace the role of the multipoles, in Fig. 5 we represent the stratified degree of randomness for the cold spots in Fig.3 but when the first 20 multipoles are removed in the WMAP's map. One can notice the general increase in randomness.

\section{Conclusions}

We obtained the sky map of CMB temperature randomness measured by Kolmogorov stochasticity parameter. 
The following conclusions can be drawn from the analysis using WMAP's 5-year data. 

First, the Galactic disk appears clearly outlined by an almost uniform $\Phi\approx 1$ coverage, which can be due to the non-Gaussian CDF there.  This marks the sensitivity of Kolmogorov's parameter to the differences between the CMB and non-cosmological radiation components.   

Second, the stability (invariance) with respect to the split regions can assign a definite degree of randomness to each of the hemispheres: 0.53, northern; 0.61, southern.  It is not clear to what extent this discrepancy can be considered genuine (cf. Copi et al 2008; Hansen et al 2008). 

Third, the randomness of the CS and of another negative temperature region in the northern hemisphere, among a sample of 20 negative temperature regions, reveals definite similarities in their stratification and agrees with what is expected for the voids. It is also notable that simulated Gaussian maps (with and without WMAP's noise) are shown to possess different behaviors and a lower degree of randomness than the real CMB ones.

These are only several of the properties that emerged from the Kolmogorov sky map. If the inhomogeneous matter distribution, more specifically, the voids are associated with the effect of randomization \cite{GK_KSP}, then K-sky maps can be unique clues to the large-scale Universe in cross-correlated studies. Both a correlation between the CS and radio surveys (radio brightness and NRAO Sky Survey number counts) \cite{Ru} and its absence \cite{Sm} has been reported. The increase in the angular resolution of the data in future experiments, notably by the Planck mission, can be particularly decisive for K-map's efficiency.   

The importance of studying of measurable properties of the inhomogeneities, in this case via K-maps, can also be related to the problem of the nature of dark energy if the latter is due to cumulative effect of inhomogeneities \cite{W,M,L,CFZ}.

In general, Kolmogorov maps can reflect underlying genuine properties for a variety of effects, cosmological and non-cosmological, and thus can open new possibilities, for example in separating signals, in analysing the information provided by conventional temperature and polarization maps. 

{\it Acknowledgments.} We thank the referee for valuable comments. The support by INTAS is acknowledged.

\end{document}